\documentclass[twocolumn,showpacs]{revtex4}
\usepackage{graphicx,epsf}
\usepackage{color}
\usepackage{bm}
\usepackage{dcolumn}
\usepackage{amsfonts}
\usepackage{mathrsfs}

\usepackage{subfig}

\begin{document}

\title{Effects of plasma nonuniformity on zero frequency zonal  structure generation by drift  Alfv\'en wave instabilities in toroidal plasmas}

\author{Zhiyong Qiu$^{1,2}$, Guangyu Wei$^3$,  Liu Chen$^{2,3,4}$,    and Ruirui Ma$^{2,5}$}

\affiliation{$^1$ Key laboratory of frontier physics in controlled nuclear fusion and  Institute of Plasma Physics, Chinese Academy of Sciences, Hefei 230031, P.R.C\\
$^2$ Center for Nonlinear Plasma Science and ENEA,   C. R. Frascati,  Italy\\
$^3$ Institute for Fusion Theory and Simulation, School of Physics, Zhejiang University, Hangzhou 310027, P.R.C\\
$^4$ Department of   Physics and Astronomy,  University of California, Irvine CA 92697-4575, U.S.A.\\
$^5$ Southwest Institute of Physics, P.O. Box 432, Chengdu 610041, P.R.C }

\begin{abstract}
Effects of plasma nonuniformity on zero frequency zonal structure (ZFZS)   excitation  by drift Alfv\'en wave (DAW) instabilities in toroidal plasmas are investigated using nonlinear gyrokinetic theory. The governing  equations describing    nonlinear interactions among ZFZS and DAWs are derived, with the contribution of DAWs self-beating and radial modulation accounted for on the same footing.   The obtained equations  are then used to derive the nonlinear dispersion relation, which is then  applied to   investigate ZFZS generation in several scenarios. In particular, it is found that, the condition for zonal flow excitation by kinetic ballooning mode (KBM) could be sensitive to plasma parameters, and more detailed investigation is needed to understand   KBM nonlinear saturation, crucial for bulk plasma transport in  future reactors.

\end{abstract}


\maketitle


\section{Introduction}
\label{sec:intro}

Shear Alfv\'en waves (SAWs) are fundamental electromagnetic oscillations in magnetized plasmas \cite{HAlfvenNature1942}. In magnetically confined fusion devices such as tokamaks, SAW instabilities driven unstable by energetic particles (EPs) including fusion alpha-particles could cause  significant redistribution and transport loss of EPs, which can lead to degradation of the confinement of both EPs and thermal components in future reactors \cite{AFasoliNF2007,LChenRMP2016}. Thus, in-depth understanding of SAW related physics, including linear excitation, nonlinear evolution,  and  saturation, is crucial for understanding of fusion reactor performance. Among various channels for SAW instability nonlinear saturation, nonlinear excitation of   zonal field structures (ZFS) is an important route \cite{AHasegawaPoF1979,ZLinScience1998,LChenPoP2000,MRosenbluthPRL1998,PDiamondPPCF2005,LChenPRL2012,ZQiuRMPP2023}, and has drawn research interest by both analytical theory \cite{LChenNF2001,LChenPRL2012,ZQiuNF2016,ZQiuPoP2016} and large-scale simulations \cite{YTodoNF2010,ABiancalaniPPCF2021,JChengPoP2017,GDongPoP2019,YChenPoP2018} in the past decade.  

Zonal field structures (ZFS) correspond to radial corrugations of plasma equilibrium, and  are characterized by toroidally symmetric and usually poloidally symmetric structures in toroidal plasmas. ZFS are linearly stable to expansion free energy, and can be nonlinearly excited by microscopic drift wave (DW) turbulences including drift Alfv\'en waves (DAWs), and in this process, scatter DW/DAW  into linearly stable short radial wavelength regime. Spontaneous excitation of zero-frequency zonal structure (ZFZS) by toroidal Alfv\'en eigenmode (TAE) \cite{CZChengAP1985} was initially investigated in Ref. \cite{LChenPRL2012} using modulational instability methodology, where the contribution of zonal current (ZC) and zonal flow (ZF) were accounted for on the same footing. It was found that, for typical plasma parameters, ZF generation is possible as the pure Alfv\'enic state is broken, i.e., the nonlinear Reynolds and Maxwell stresses don't exactly cancel each other,  by toroidicity; and ZC generation is preferred with a much lower threshold condition, and can be estimated by $|\delta B_r/B_0|\sim O(10^{-4})$, comparable to typical electromagnetic fluctuation level in present day tokamak experiments \cite{WHeidbrinkPRL2007}.   Here, $\delta B_r$ is the perturbed magnetic field associated with the primary TAE, and $B_0$ is the equilibrium on-axis toroidal magnetic field. The  ZFZS growth rate is approximately  proportional to the amplitude of the pump TAE as the nonlinear drive overcomes the threshold due to frequency mismatch.

It was further found that, in the presence of resonant EP drive to TAE, ZF can be  excited by TAE even when TAE amplitude is small, with the ZF growth rate being approximately  twice of instaneous TAE growth rate \cite{ZQiuPoP2016}. This thresholdless   ``forced driven" (also called ``passive excitation" in some literatures)   process was shown to be dominated  by the contribution of EPs to the curvature coupling term, and was  observed by  simulations of Alfv\'en instability nonlinear dynamics using both hybrid code \cite{YTodoNF2010} and particle-in-cell codes \cite{YChenPoP2018,ABiancalaniPPCF2021}. A unified theory containing both the ``forced driven" process and the spontaneous excitation was presented in Ref. \citenum{ZQiuNF2017}, which shows that these two processes occur as the corresponding nonlinear contribution of EPs  or thermal plasmas take over the nonlinear coupling. It is also found that, the forced-driven process can also occur as plasma nonuniformity associated with diamagnetic drift is accounted for, with the generated ZF amplitude being proportional to the local diamagnetic frequency \cite{LChenWLIS2023, QFangNF2024}.  In the following discussion,  we will term the ``forced driven" as ``beat driven" based on recent advances of understanding \cite{LChenPoP2024}. 

The above analysis and the obtained understanding, using TAE as a paradigm, can be applied to other SAW instabilities, with proper understanding of their respective properties.  For example, it is shown in Ref. \cite{ZQiuNF2016} that, ZFZS can be excited by beta-induced Alfv\'en eigenmode (BAE) \cite{WHeidbrinkPRL1993,FZoncaPPCF1996,HZhangPST2013,JChengPoP2017}, with ZF dominating due to vanishing $k_{\parallel B}$. Here, $k_{\parallel B}$ is the parallel wavenumber of BAE. The theory of Ref. \cite{ZQiuNF2016} may represent a series of low frequency Alfv\'en modes (LFAMs) \cite{FZoncaPPCF1996,LChenPoP2017,WHeidbrinkNF2021,RMaPPCF2022} that can be excited by both EPs and thermal plasmas in different spatiotemporal scales, with plasma nonuniformity expected to play important roles with their frequencies comparable to diamagnetic frequency $\omega_*$ due to plasma nonuniformity.    In this work, we will extend the theory of ZFZS generation to nonuniform plasma, motivated by the recent discovery that, system nonuniformity can both qualitatively and quantitatively affect the parametric decay of kinetic Alfv\'en waves (KAW) due to the diamagnetic effects \cite{LChenPoP2022}.   Here, for nonuniformity, we mean the plasma profile associated with $\omega_*$, as the previous analysis   labelled ``uniform", is intrinsically nonuniform due to, e.g., toroidicity of magnetic geometry.  The following analysis,  using general WKB representation, can be applied to DAWs of a  broad frequency range, from TAE down to  kinetic ballooning mode (KBM)  frequency range   \cite{FZoncaPoP1999,JKimPoFB1993}, and is expected to be crucial for   KBM with 
 $\omega\sim\omega_*$ and higher toroidal mode numbers with $k_{\perp}\rho_i\lesssim O(1)$.  Here, $\omega$ is the mode frequency,  $\omega_*$ is the characteristic ion diamagnetic frequency due to profile nonuniformity, $k_{\perp}$ is the perpendicular wavenumber, and $\rho_i$ is the ion Larmor radius. The present work, is thus, of particular importance to future reactor scale tokamaks with thermal to magnetic pressure ratio significantly higher than present day machines, and KBM  is expected to be crucial for thermal plasma transport, while there is still ongoing debate on effects of ZFS in saturating KBM \cite{GDongPoP2019,AIshizawaPoP2019,GRenNF2022}.  The analysis, however,  is  general, and can be applied to other DAW instabilities, e.g., TAE, though the effects of system nonuniformity  are expected to be less important for TAE, in that $|\omega_*/\omega| \ll 1$ for typical scenarios.

The rest of the manuscript is organized as follows. In Sec. \ref{sec:model}, the theoretical model and governing equations are introduced, which are then used in Sec. \ref{sec:zfzs_generation} to derive the modulational dispersion relation for ZFZS generation by DAWs. The obtained nonlinear dispersion relation, is applied to study the effects of plasma nonuniformity on ZFZS generation in Sec. \ref{sec:application}. Finally, summary and discussion are presented in Sec. \ref{sec:summary}.

\section{Theoretical model}
\label{sec:model}

For the nonlinear excitation of ZFZS by DAW instabilities in nonuniform plasmas, the analysis follows closely that of Ref. \cite{LChenPRL2012}, using the methodology of modulational instability. The ballooning representation  is adopted for the pump DAW and its lower/upper sidebands due to ZFZS modulation:
\begin{eqnarray}
\delta\phi_0 &=& A_0 e^{i(n_0\phi-\hat{m}_0\theta-\omega_0 t)}\sum_j e^{-ij\theta}\Phi_0(x-j)+c.c.,\nonumber\\
\delta\phi_{\pm} &=& A_{\pm} e^{\pm i(n_0\phi-\hat{m}_0\theta-\omega_0 t)} e^{i(k_Z r-\omega_Zt)}\nonumber\\
&&\times\sum_j e^{\mp ij\theta}\Phi_0(x-j)+c.c.,\nonumber
\end{eqnarray}
while ZFZS   potential can be taken as
\begin{eqnarray}
\delta\phi_Z=A_Z e^{i(k_Z r-\omega_Z t)}.\nonumber
\end{eqnarray}
In the above expressions, $A$ is the mode envelope amplitude, $n$ is the toroidal mode number, $\hat{m}$ is the reference poloidal mode number, $m=\hat{m}+j$ is the poloidal mode number, $\Phi_0(x-j)$ is the parallel mode structure, subscripts ``$0$", ``$\pm$" and ``$Z$" represent pump DAW, its upper/lower sidebands and ZFZS,  and frequency/wave number matching conditions for nonlinear mode coupling are applied.

The analysis follows closely the standard modulational instability approach, where the nonlinear ZFZS and DAW sidebands equations are derived, which then couple and yield the nonlinear modulational instability dispersion relation for ZFZS excitation. While ``modulational instability'' is used, the process can go beyond spontaneous excitation by modulational instability, and beat-driven process can also be included in the general equations \cite{ZQiuNF2017,LChenPoP2024}. The perturbed distribution function, $\delta f_s$ with $s=e,i$, obeys the nonlinear gyrokinetic equation \cite{EFriemanPoF1982}:
\begin{eqnarray}
\delta f_s=-\left(\frac{e}{T}\right)_s\delta\phi F_{0s} +\exp(-\bm{\rho}_s\cdot\nabla)\delta H_s,
\end{eqnarray}
with the nonadiabatic particle response $\delta H_s$ derived from nonlinear gyrokinetic equation \cite{EFriemanPoF1982}
\begin{eqnarray}
\left(\partial_t +  v_{\parallel}\partial_l  +\mathbf{v}_d\cdot\nabla\right)\delta H_s&=&-i \left(\frac{e}{T}\right)_s \left(\omega-\omega^t_*\right) F_{0s}  J_k\delta L_k\nonumber\\
&&- \Lambda^{k'}_{k''}J_{k'}\delta L_{k'}\delta H_{k''}.  \label{eq:gk}
\end{eqnarray}
Here,  $l$ is the arc length along the equilibrium magnetic field line,  $\mathbf{v}_d\equiv \mathbf{\hat{b}}\times[(v^2_{\perp}/2)\nabla\ln B_0 + v^2_{\parallel}\mathbf{\hat{b}}\cdot\nabla\mathbf{\hat{b}}]$ is the magnetic drift velocity, $\mathbf{\hat{b}}\equiv\mathbf{B}_0/B_0$ is  the unit vector along the equilibrium magnetic field line, $J_k\equiv J_0(k_{\perp}\rho_i)$ is Bessel function of zero index accounting for finite Larmor radius (FLR) effects, $\delta L_k\equiv \delta\phi-(k_{\parallel}v_{\parallel}/\omega)\delta\psi$ with $\delta\psi\equiv\omega\delta A_{\parallel}/(ck_{\parallel})$ and  $\delta\psi=\delta\phi$ corresponding to vanishing parallel electric field $\delta E_{\parallel}$ in the ideal MHD condition, $\omega^t_*=\omega_*[1+\eta (v^2/v^2_{th}-3/2)]$ is the diamagnetic drift frequency associated with plasma nonuniformity with $\omega_*=-ck_{\theta}T/(q B L_n)$, $\eta=L_n/L_T$,  $L_n$ and $L_T$ are the scale lengths of density and temperature nonuniformity, respectively, and   $ \Lambda^{k'}_{k''}\equiv (c/B_0)\mathbf{\hat{b}}\cdot\mathbf{k''}\times\mathbf{k'}$  denotes the perpendicular nonlinearity with   the matching condition $\mathbf{k}=\mathbf{k}'+\mathbf{k}''$.

The governing field equations, in the $\beta\ll1$ limit with magnetic compression being negligible,  can be derived from  the quasi-neutrality condition
\begin{eqnarray}
\frac{n_0e^2}{T_i}\left(1+\frac{T_i}{T_e}\right)=\sum_{s=e,i}\left\langle e_s J_k\delta H_k\right\rangle,\label{eq:qn}
\end{eqnarray}
and nonlinear gyrokinetic vorticity equation \cite{LChenNF2001,LChenJGR1991}
\begin{eqnarray}
&&\frac{c^2}{4\pi \omega^2_k}B\frac{\partial}{\partial l}\frac{k^2_{\perp}}{B}\frac{\partial}{\partial l}\delta \psi_k +\frac{n_0e^2}{T_i}\left(1-\frac{\omega_{*i}}{\omega}\right)(1-\Gamma_0)\delta\phi_k\nonumber\\
&&-\frac{e^2}{T_i}\frac{\omega_{*i}}{\omega}\eta_i\left\langle F_0 J^2_0 \left(\frac{v^2}{v^2_{th}}-\frac{3}{2}\right)\right\rangle\delta\phi_k\nonumber\\
&&-\sum_{s=e,i}\left\langle\frac{q}{\omega_k}J_k\omega_d\delta H_k \right\rangle_s\nonumber\\
&=&-\frac{i}{\omega_k} \Lambda^{k'}_{k''}\left [ \frac{c^2}{4\pi}k''^2_{\perp} \frac{\partial_l\delta\psi_{k'}\partial_l\delta\psi_{k''}}{\omega_{k'}\omega_{k''}} \right.\nonumber\\
&&\left.+ \left\langle e(J_kJ_{k'}-J_{k''})\delta L_{k'}\delta H_{k''}\right\rangle \right].
\label{eq:vorticityequation}
\end{eqnarray}
The terms on the left hand side of equation (\ref{eq:vorticityequation}) are the field line bending, inertial, and curvature coupling terms, with the SAW dispersion relation being straightforwardly obtained from the balance of the former two, while the curvature coupling term plays crucial role in the low frequency SAW spectrum including KBM  \cite{JKimPoFB1993,FZoncaPoP1999}. The terms on the right hand side correspond to nonlinear Maxwell stress (MX)  and gyrokinetic Reynolds stress (RS), respectively.
In the following analysis, electron force balance equation, i.e., nonlinear Ohm's law, is sometimes used alternatively to simplify the analysis.

\section{General equation for ZFZS generation}
\label{sec:zfzs_generation}

The zonal current equation can be derived from the parallel component of the electron force balance equation,
\begin{eqnarray}
\delta E_{\parallel}+\mathbf{\hat{b}}\cdot\mathbf{\delta u}_{\perp}\times\mathbf{\delta B}_{\perp}=0,
\end{eqnarray}
describing ZC generation due to dynamo effects.  Here, noting $\mathbf{\delta u}_{\perp}\equiv \nabla\delta\phi\times\mathbf{\hat{b}}/B_0$ and $\mathbf{\delta B}_{\perp}=\nabla\times\delta A_{\parallel}\mathbf{\hat{b}}$, and that for ZFZS with $n/m=0/0$ and thus $\delta E_{\parallel Z}=-\partial_t\delta A_{\parallel Z}$, one has, for ZC generation due to pump DAW and its upper/lower sidebands coupling,
\begin{eqnarray}
i\omega_Z\delta A_{\parallel Z}&=&\frac{c}{B_0}k_Zk_{\theta 0}k_{\parallel 0} \left[\left(\frac{\sigma_{* -}}{\omega_0-\omega_Z}-\frac{\sigma_{* 0}}{\omega_0}\right)\delta\phi_0\delta\phi_- \right.\nonumber\\ &&\left.-\left(\frac{\sigma_{* +}}{\omega_0+\omega_Z}-\frac{\sigma_{* 0}}{\omega_0}\right)\delta\phi_{0^*}\delta\phi_+ \right],
\end{eqnarray}
with the expression of $\sigma_{*k}=\delta\psi_k/\delta\phi_k$ given by
\begin{eqnarray}
\sigma_{*k}\equiv \frac{1+\tau-\tau\Gamma_k +\tau \langle (F_0/n_0)J^2_0(\omega^t_{*i}/\omega)\rangle_k}{(1-\omega_{*e}/\omega)_k},
\end{eqnarray}
and  again, $\sigma_{*k}=1$ corresponding to ideal MHD condition.

Noting $\sigma_{*\pm}=\sigma_{*0}(\omega_0\pm\omega_Z,k_Z)$, one has
\begin{eqnarray}
\frac{\sigma_{*\pm}}{\omega_0\pm\omega_Z}-\frac{\sigma_{*0}}{\omega_0}=\pm\frac{\partial}{\partial\omega_0}\left(\frac{\sigma_{*0}}{\omega_0}\right)\omega_Z + \frac{1}{2}\frac{\partial^2}{\partial k^2_r}\left(\frac{\sigma_{*0}}{\omega_0}\right) k^2_Z.
\end{eqnarray}
Defining $\delta A_{\parallel Z}=ck_{\parallel 0}\delta\psi_Z/\omega_0$ using the frequency and parallel wavenumber of the pump DAW, one has
\begin{eqnarray}
\delta\psi_Z&=&\frac{i}{B}k_Zk_{\theta 0}\omega_0 \left[\frac{\partial}{\partial\omega_0}\left(\frac{\sigma_{*0}}{\omega_0}\right) (\delta\phi_-\delta\phi_0 + \delta\phi_+\delta\phi_{0^*})\right.\nonumber\\
&&\left. - \frac{1}{2\omega_Z}\frac{\partial^2}{\partial k^2_r}\left(\frac{\sigma_{*0}}{\omega_0}\right) k^2_Z (\delta\phi_-\delta\phi_0 - \delta\phi_+\delta\phi_{0^*})  \right]. \label{eq:ZC}
\end{eqnarray}
The first term on the right hand side  of equation (\ref{eq:ZC}) comes from the frequency difference between the DAW sideband to that of the pump, and corresponds to the ``beat-driven" of ZC that has been extensively studied \cite{LChenPRL2012}. The second term, on the other hand, comes from plasma nonuniformity and kinetic effects, will contribute to ``spontaneous excitation" of ZC, and is not accounted for in previous publications, focusing on TAE frequency where   $\omega_*$ effects can be neglected due to the $|\omega_*/\omega|\ll1$ ordering typical of TAE \cite{LChenPRL2012,LChenNF2022}.

The ZF equation can be derived from the nonlinear vorticity equation, by substituting the particle responses to DAWs  into RS, and one obtains:
\begin{eqnarray}
\chi_{iZ}\delta\phi_Z = - i\frac{ck_Zk_{\theta0}}{B_0 \omega_Z} \left(\alpha_-\delta\phi_0\delta\phi_--\alpha_+\delta\phi_{0^*}\delta\phi_+\right),\label{eq:ZF}
\end{eqnarray}
with
\begin{eqnarray}
\chi_{iZ}\equiv 1-\Gamma_Z  -\frac{T_i}{n_0e^2\delta\phi_Z} \sum_{s=e,i}\left\langle \overline{\frac{q}{\omega}J_Z\omega_d\delta H_Z}\right\rangle
\end{eqnarray}
being the neoclassical inertial enhancement dominated by trapped particle contribution  \cite{MRosenbluthPRL1998}, and
\begin{eqnarray}
\alpha_{\pm}=&&\left\langle (J_ZJ_{\pm}J_0-J^2_{\pm})\frac{F_0}{n_0}\left(1-\frac{\omega^t_{*i}}{\omega}\right)_{\pm}\right\rangle \nonumber\\
&-&\left\langle (J_ZJ_{\pm}J_0-J^2_0)\frac{F_0}{n_0}\left(1-\frac{\omega^t_{*i}}{\omega}\right)_0\right\rangle\nonumber\\
 &-&\frac{c^2}{4\pi} (k^2_{\perp\pm}-k^2_{\perp0})\frac{k_{\parallel\pm}k_{\parallel0}\sigma_{*0}\sigma_{*\pm}}{\omega_0\omega_{\pm}}.
\end{eqnarray}
 In deriving equation (\ref{eq:ZF}),  the linear particle response to DAW   has been used \cite{LChenNF2022}, i.e., 
\begin{eqnarray}
\delta H^L_{Ai}&=& \frac{e}{T_i}\left(1-\frac{\omega^t_{*i}}{\omega}\right)_A J_A F_{Mi}\delta\phi_A,\nonumber\\
\delta H^L_{Ae}&=&-\frac{e}{T_e}\left(1-\frac{\omega^t_{*e}}{\omega}\right)_A F_{Me}\delta\psi_A.\nonumber
\end{eqnarray}
Here, the superscript ``L" denotes linear response, and the subscript ``A" denotes DAW, which can be used for both the pump DAW and its lower/upper sidebands. 

In the long wavelength $|k_{\perp}\rho_i|\ll1$ and uniform plasma $|\omega^t_{*i}/\omega|\ll1$ limit, equation (\ref{eq:ZF}) recovers the results of Ref. \citenum{LChenPRL2012}, with $\alpha_{\pm}$ denoting  MX and RS  competition, and $\chi_{iZ}\simeq 1.6 b_{Z} q^2/\sqrt{\epsilon}$ with $b_Z\equiv k^2_Z\rho^2_i$.

The governing equations describing the DAW upper sideband generation due to pump DAW and ZFZS coupling, can be derived from the quasi-neutrality condition and the nonlinear vorticity equation. The governing equations for the DAW  lower sideband generation, can be derived similarly. 
The nonlinear electron response to $\delta\phi_+$, can be derived from the nonlinear gyrokinetic equation. Noting the  $k_{\parallel+}v_e\gg\omega_+$ ordering, one has,
\begin{eqnarray}
\delta H^{NL}_{e +}\simeq  i \frac{ ck_Zk_{\theta0}}{B_0\omega_0}\frac{e}{T_e} F_0\delta\psi_0\left[\delta\phi_Z -\left(1- \frac{\omega^t_{*e}}{\omega} \right)_0   \delta\psi_Z  \right],
\end{eqnarray}
with the superscript ``NL" denoting nonlinear particle response. 
The nonlinear ion response to $\delta\phi_+$, on the other hand,  can be derived noting the $\omega\sim\omega_*\gg k_{\parallel}v_i, \omega_d$ ordering, and one obtains
\begin{eqnarray}
\delta H^{NL}_{i +}\simeq -i\frac{ck_Zk_{\theta 0}}{B_0\omega_+}  J_0J_Z\frac{e}{T} F_0 \left(\frac{\omega^t_{*i}}{\omega}\right)_0 \delta\phi_0\delta\phi_Z.
\end{eqnarray}
Substituting the derived particle responses  into the quasi-neutrality condition, one obtains
\begin{eqnarray}
\delta\psi_+ &=& \sigma_{*+}\delta\phi_+\nonumber\\
 &+&\frac{ i(c/B_0)k_Zk_{\theta 0}}{\omega_0(1-\omega_{*e}/\omega)_+} \delta\phi_0 \left[ -\sigma_{*0}\left( 1- \frac{\omega_{*e}}{\omega}\right)_0 \delta\psi_Z\right.\nonumber\\
&+& \left.\left(\sigma_{*0}+\tau\left\langle \frac{F_0}{n_0} J_0J_ZJ_+\left(\frac{\omega^t_{*i}}{\omega}\right)_0\right\rangle\right)\delta\phi_Z  \right]. \label{eq:sideband_qn}
\end{eqnarray}

The other equation can be derived from vorticity equation. Substituting the linear ion  response  to ZFZS and pump DAW  into RS,  we have
\small\begin{eqnarray}
&&b_+\left[-\frac{k^2_{\parallel +}V^2_A}{\omega^2_+}\delta\psi_+   + \frac{1-\Gamma_+}{b_+}\left(1-\frac{\omega_{*i}}{\omega}\right)_+ \delta\phi_+\right. \nonumber\\
&-&\left.\frac{\eta_i\omega_{*i}}{\omega_+ b_+} \left\langle \frac{F_0}{n_0} J^2_+\left(\frac{v^2}{v^2_{th}}-\frac{3}{2}\right)\right\rangle\delta\phi_+ 
+\frac{T_i}{e b_+}  \left\langle \frac{F_0}{n_0} J_+\frac{\omega_d}{\omega_+}\delta H^L_{+i}\right\rangle \right] \nonumber\\
&\simeq&-i\frac{c}{B_0\omega_+}k_Zk_{\theta0} \left\{-(b_Z-b_0)\frac{k^2_{\parallel0}V^2_A}{\omega^2_0}\delta\phi_0\delta\psi_Z\right.\nonumber\\
&+&\left.\left[ \Gamma_0-\Gamma_Z+\left\langle (J_+J_ZJ_0 - J^2_0)\left(\frac{\omega^t_{*i}}{\omega}\right)_0\frac{F_0}{n_0}\right\rangle \right]\delta\phi_Z\delta\phi_0\right\}.\nonumber\\
\label{eq:sideband_vorticity}
\end{eqnarray}\normalsize

Substituting equation (\ref{eq:sideband_qn}) into (\ref{eq:sideband_vorticity}), one obtains, the equation describing $\delta\phi_+$ evolution due to $\delta\phi_0$ and $\delta\phi_Z$ coupling
\begin{eqnarray}
&&b_+\epsilon_{A+} \delta\phi_+  \simeq -i\frac{c k_Zk_{\theta0} }{B_0\omega_+} \left( \beta_+  \delta\phi_Z  + \gamma_+ \delta\psi_Z   \right)\delta\phi_0, \label{eq:upper_sideband}
\end{eqnarray}
with
\begin{eqnarray}
&&\epsilon_{A+}  \equiv  -\frac{k^2_{\parallel +}V^2_A}{\omega^2_+}\sigma_{*+}  +\frac{T_i}{e b_+\delta\phi_+}  \left\langle \frac{F_0}{n_0} J_+\frac{\omega_d}{\omega_+}\delta H^L_{+i}\right\rangle     \nonumber\\
& & +  \frac{1-\Gamma_+}{b_+}\left(1-\frac{\omega_{*i}}{\omega}\right)_+ + \frac{\eta_i\omega_{*i}}{\omega_+ b_+} \left\langle \frac{F_0}{n_0} J^2_+\left(\frac{v^2}{v^2_{th}}-\frac{3}{2}\right)\right\rangle \nonumber
\end{eqnarray}
being the DAW  upper sideband dispersion relation in the WKB limit,
\begin{eqnarray}
\beta_+&=&  \Gamma_0-\Gamma_Z+\left\langle (J_+J_ZJ_0 - J^2_0)\left(\frac{\omega^t_{*i}}{\omega}\right)_0\frac{F_0}{n_0}\right\rangle\nonumber\\  &-&\frac{k^2_{\parallel +}V^2_A b_+}{\omega_+(\omega -\omega_{*e})_+} \left( \sigma_{*0}+\tau \left\langle J_+J_0J_Z\left(\frac{\omega^t_{*i}}{\omega}\right)_0\frac{F_0}{n_0}\right\rangle\right),\nonumber
\end{eqnarray}
and
\begin{eqnarray}
\gamma_+\equiv   -\left[\frac{k^2_{\parallel0}V^2_A}{\omega^2_0}(b_Z-b_0)-\frac{k^2_{\parallel0}V^2_A}{\omega^2_+}b_+\right]\sigma_{*0}.
\end{eqnarray}
Here, the terms proportional to $b_+$  are from the parallel electric field through equation (\ref{eq:sideband_qn}), while the rest are  from the generalized gyrokinetic RS and MX.

The lower sideband equation can be derived similarly as
\begin{eqnarray}
b_- \epsilon_{A-}\delta\phi_-&=&-i\frac{ck_Zk_{\theta0}}{B_0|\omega_-|}  \left(\beta_- \delta\phi_Z+\gamma_-\delta\psi_Z\right)\delta\phi_{0^*},\label{eq:lower_sideband}
\end{eqnarray}
with $\beta_-$ and $\gamma_-$ defined similarly to $\beta_+$ and $\gamma_+$.

\section{Modulational dispersion relation for ZFZS generation by DAW}
\label{sec:application}

The nonlinear  dispersion relation for   ZFZS excitation by DAW, can be derived from equations (\ref{eq:ZC}), (\ref{eq:ZF}), (\ref{eq:upper_sideband}) and (\ref{eq:lower_sideband}).  Noting the major  difference between $\beta_+$ and $\beta_-$, and $\gamma_+$ and $\gamma_-$  are multiplied by a $k^2_{\parallel0}$ factor, which vanishes for low frequency range where effects associated with plasma nonuniformities could be important, we can safely take $\hat{\gamma}\equiv \gamma_+=\gamma_-$ and $\hat{\beta}\equiv\beta_+=\beta_-$ from now on.    Taking $(\ref{eq:ZF})\times \hat{\beta}/\chi_{iZ} + (\ref{eq:ZC}))\times \hat{\gamma}$, and substituting $\delta\phi_{\pm}$ from equations (\ref{eq:upper_sideband}) and (\ref{eq:lower_sideband}), one obtains
\begin{eqnarray}
&&\left(\frac{c}{B_0}k_Zk_{\theta0}\right)^2\frac{|\delta\phi_0|^2}{b_+}\left[\hat{\gamma}\frac{\partial}{\partial\omega_0}\left(\frac{\sigma_{*0}}{\omega_0}\right)\left(\frac{1}{\epsilon_{A-}}+\frac{1}{\epsilon_{A+}}\right)\right.\nonumber\\
& +&\left. \frac{\hat{\gamma}k^2_Z }{2\omega_Z}\frac{\partial^2}{\partial k^2_r}\left(\frac{\sigma_{*0}}{\omega_0}\right) \left(\frac{1}{\epsilon_{A+}}-\frac{1}{\epsilon_{A-}}\right)\right.\nonumber\\
& +&\left. \frac{\hat{\beta}}{\omega_Z\omega_0\chi_{iZ}} \left( \frac{\alpha_+}{\epsilon_{A+}}-\frac{\alpha_-}{\epsilon_{A-}}\right) \right]=1.\label{eq:modulational_DR}
\end{eqnarray}
Equation (\ref{eq:modulational_DR}) is the desired modulational dispersion relation for ZFZS generation by DAW, with the effects of system nonuniformity and kinetic effects such as finite orbit width and FLR   effects  systematically accounted for. It also includes the contribution of beat-driven and spontaneous excitation on the same footing, which is reflected in, e.g., equation (\ref{eq:ZC}) for zonal current and equation (\ref{eq:ZF}) for zonal flow  noting the dependence of ``$\alpha_{\pm}$" on $\omega_Z$ and $k_Z$.    Equation (\ref{eq:modulational_DR}) is general and needs numerical solution for  proper understanding of the ZFZS generation by DAW. It can, however, be analyzed in various limits, by using of the knowledge from previous works.

\subsection{Uniform plasma  limit with $k^2_{\parallel 0}V^2_A/\omega^2_0\simeq 1$}

We   start from the TAE frequency mode in the  uniform plasma ($\omega_*=0$)    ideal MHD ($\sigma_{*0}=1$) limit.  Noting $|\omega_Z|\ll |\omega_0|$ and $k^2_{\parallel0}V^2_A\simeq \omega^2_0$ to the leading order,  one has $\alpha_+=\alpha_-\simeq b_Z (1-k^2_{\parallel0}V^2_A/\omega^2_0)$, $\hat{\gamma}\simeq 2b_0$, and $\hat{\beta}\simeq -2 b_0$.  We further note $\epsilon_{A\pm}\equiv (\partial\epsilon_{A0}/\partial\omega_0) (\pm \omega_Z+\Delta_T)$, with  $\Delta_T\equiv (\partial^2\epsilon_{A0}/\partial k^2_r) k^2_Z/(2\partial\epsilon_{A0}/\partial\omega_0)$ being frequency mismatch and $\omega_Z=i\gamma_Z$ with $\gamma_Z$ being the nonlinear growth rate of ZFZS, we then  obtain
\begin{eqnarray}
\gamma^2_Z=-\Delta^2_T&+&\left(\frac{c}{B_0}k_Zk_{\theta0}\right)^2\frac{2 b_0 |A_0|^2}{b_+\omega_0(\partial\epsilon_{A0}/\partial\omega_0)} \nonumber\\
&\times&\left[\frac{\Delta_T}{\omega_0} + \frac{4(1-k^2_{\parallel0}V^2_A/\omega^2_0)}{\hat{\chi}_{iZ}}\right]. \label{eq:DR_uniform}
\end{eqnarray}
Equation (\ref{eq:DR_uniform}) recovers the result of Ref. \cite{LChenPRL2012},  describing ZFZS generation as the nonlinear drive overcomes the threshold due to frequency mismatch, with the two terms in the square brackets representing the contribution from ZC and ZF, respectively.  It can be seen that, the contribution of ZF is limited by two factors, i.e., neoclassical shielding ($\hat{\chi}_{iZ}\equiv \chi_{iZ}/b_Z\simeq 1.6 q^2/\sqrt{\epsilon} \gg1$) as well as RS \& MX cancellation with $1-k^2_{\parallel0}V^2_A/\omega^2_0\sim O(\epsilon)$, rendering  ZC generation   preferred with much lower threshold for modes in the TAE frequency range with $\omega^2_0\simeq k^2_{\parallel0}V^2_A$ if  $\Delta_T/\omega_0>0$.  On the other hand, if $\Delta_T/\omega_0<0$, ZFZS generation is still possible, given the pump TAE is excited in the upper half of the toroidicity induced SAW continuum gap, and the frequency mismatch is small enough. It, however, requires a relatively higher pump TAE amplitude to overcome the threshold due to frequency mismatch. The  physics picture is extensively discussed in Ref. \cite{LChenPRL2012} on ZFZS excitation by TAE, and interested readers may refer to the original publication for more details. 

\subsection{Nonuniform plasma with predominant  ZC generation}

If we consider the general case with $\hat{\chi}_{iZ}\gg1$  and ZC generation (proportional to $\hat{\gamma}$ in equation (\ref{eq:modulational_DR}))    is dominant,  the modulational dispersion relation can be written as
\begin{eqnarray}
\gamma^2_Z&=& -\Delta^2_T  +  2 \left(\frac{c}{B_0}k_Zk_{\theta0}\right)^2 \frac{\hat{\gamma}|A_0|^2}{b_+(\partial\epsilon_{A0}/\partial\omega_0)}\nonumber\\
&\times& \left[\frac{\partial}{\partial\omega_0}\left(\frac{\sigma_{*0}}{\omega_0}\right)\Delta_T -  \frac{1}{2}\frac{\partial^2}{\partial k^2_r} \left(\frac{\sigma_{*0}}{\omega_0}\right) k^2_Z  \right]. 
\end{eqnarray}
Noting the expression of $\epsilon_A$, one has,
\begin{eqnarray}
\frac{\partial^2}{\partial k^2_r} \left(\frac{\sigma_{*0}}{\omega_0}\right) &\simeq& -\frac{\omega_0}{k^2_{\parallel0}V^2_A} \frac{\partial^2\epsilon_{A0}}{\partial k^2_r}, \nonumber\\
\frac{\partial}{\partial\omega_0}\left(\frac{\sigma_{*0}}{\omega_0}\right) &\simeq& \frac{\sigma_{*0}}{\omega^2_0} -\frac{\omega_0}{k^2_{\parallel0}V^2_A}  \frac{\partial\epsilon_{A0}}{\partial\omega_0}.\nonumber
\end{eqnarray}
On the other hand, $\hat{\gamma}\simeq 2b_0\sigma_{*0}$,   and we have
\begin{eqnarray}
\gamma^2_Z=-\Delta^2_T &+& 4 \left(\frac{c}{B_0}k_Zk_{\theta0}\right)^2 \frac{b_0}{b_+}  \frac{\Delta_T \sigma^2_{*0}|A_0|^2}{\omega^2_0(\partial\epsilon_{A0}/\partial\omega_0)},
\end{eqnarray}
i.e., compared to the uniform plasma case (neglecting contribution of $\omega_*$) with predominant ZC generation, the nonlinear drive is quantitatively enhanced by $\sigma^2_{*0}$, while the qualitative picture is not altered.

\subsection{ZF generation by DAW with $k^2_{\parallel0}V^2_A\ll\omega^2_0$}

An important parameter regime for plasma nonuniformity to play crucial role to SAW instability is the low frequency range with mode frequency comparable to  the diamagnetic  and/or ion transit frequency range \cite{FZoncaPPCF1996,LChenPoP2017,RMaPPCF2022}, where the modes are characterized by $\omega^2_0\gg k^2_{\parallel0}V^2_A$. It is shown for the BAE case that, ZF generation dominates due to vanishing $k^2_{\parallel}V^2_A/\omega^2$ such that the pure SAW state is broken due to  RS dominance over MX \cite{ZQiuNF2016}.  To analyze this parameter regime, we take $k_{\parallel0}=0$ limit in equation (\ref{eq:modulational_DR}) and focus on the effects associated with plasma nonuniformity.  In this parameter regime, we have  $\hat{\gamma}=0$, 
\begin{eqnarray}
\beta_{\pm}&=&\Gamma_0-\Gamma_Z+\left\langle \frac{F_0}{n_0} \left(J_{\pm}J_0J_Z-J^2_0\right)\left(\frac{\omega^t_{*i}}{\omega}\right)_0\right\rangle,\\
\alpha_{\pm}&=&\left\langle \left(J_ZJ_0J_{\pm}-J^2_{\pm}\right)\frac{F_0}{n_0}\left(1-\frac{\omega^t_{*i}}{\omega}\right)_{\pm}\right\rangle\nonumber\\
&&-\left\langle \left(J_ZJ_0J_{\pm}-J^2_0\right)\frac{F_0}{n_0}\left(1-\frac{\omega^t_{*i}}{\omega}\right)_0\right\rangle,
\end{eqnarray}
the modulational dispersion relation, equation (\ref{eq:modulational_DR}), then becomes:
\begin{eqnarray}
\left(\frac{c}{B_0}k_Zk_{\theta0}\right)\frac{|A_0|^2\hat{\beta}}{b_+\omega_Z\omega_0\chi_{iZ}}\left[\frac{\alpha_+}{\epsilon_{A+}}-\frac{\alpha_-}{\epsilon_{A-}}\right]=1.
\end{eqnarray}
Noting again, $\epsilon_{\pm}=(\partial\epsilon_{A0}/\partial\omega_0)(\pm\omega_Z+\Delta)$, and 
\begin{eqnarray}
\alpha_++\alpha_-&\simeq&\frac{2\omega_Z}{\omega_0}\left\langle \left(J_ZJ_0J_Z-J^2_+\right)\frac{F_0}{n_0}\frac{\omega^t_{*i0}}{\omega_0}\right\rangle,\\
\alpha_+-\alpha_-&\simeq&-2\left\langle \left(J^2_+-J^2_0\right)\frac{F_0}{n_0}\left(1-\frac{\omega^t_{*i}}{\omega}\right)_0\right\rangle, 
\end{eqnarray}
we then have
\begin{eqnarray}
\gamma^2_Z&=&-\Delta^2 +\left(\frac{c}{B_0}k_Zk_{\theta}\right)^2\frac{|A_0|^2\hat{\beta}}{b_+\omega_0\chi_{iZ}}\frac{2}{\partial\epsilon_{A0}/\partial\omega_0}\nonumber\\
&\times& \left[-i\frac{\Delta_T}{\gamma_Z}\left\langle (J_+J_ZJ_0-J^2_+)\left(\frac{\omega^t_{*i}}{\omega}\right)_0\frac{F_0}{n_0}\right\rangle\right.\nonumber\\
&&\left. +\left\langle (J^2_+-J^2_0) \left(1-\frac{\omega^t_{*i}}{\omega}\right)_0\frac{F_0}{n_0}\right\rangle \right].\label{eq:ZF_KBM}
\end{eqnarray}

\begin{figure}[htbp!]
  \centering
  \begin{minipage}{.99\linewidth}
      \centering
      \includegraphics[width=.9\linewidth]{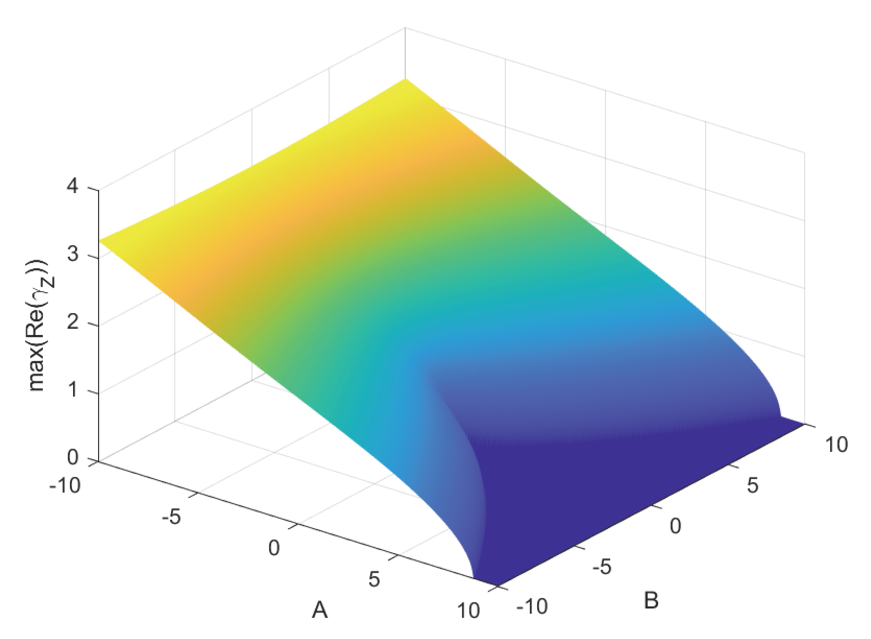}
  \end{minipage}

  \begin{minipage}{.99\linewidth}
      \centering
      \includegraphics[width=.9\linewidth]{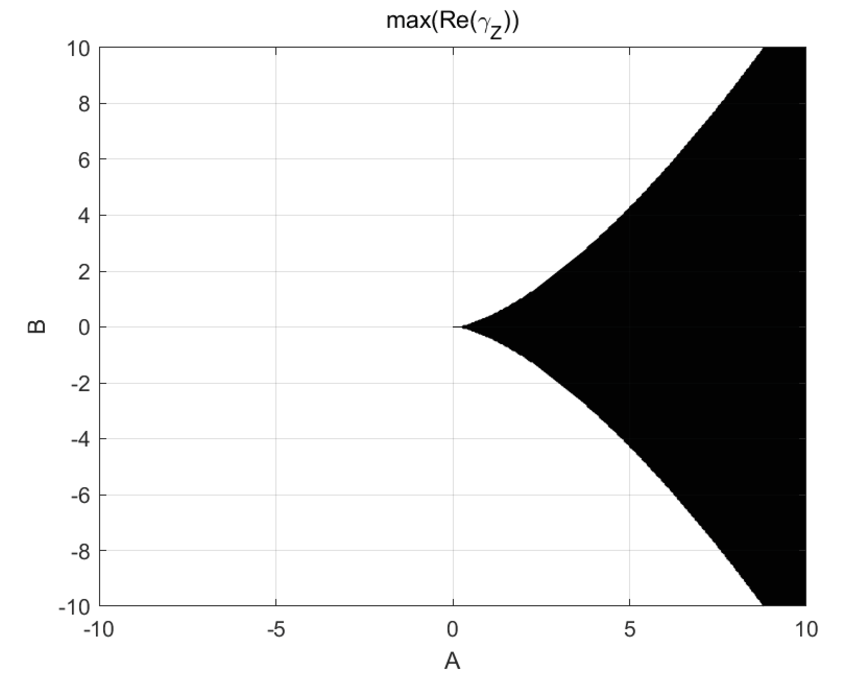}
  \end{minipage}
  \caption{Dependence of ZFZS growth rate on parameters $A$ and $B$. Here, the maximum value of the real part of $\gamma_Z$ is plotted. In Fig. 1b, the  region in black corresponds to ZFZS marginally stable, while the   region in white correspond to ZFZS unstable. The boundary between the white and back regions is determined by $B^2/4 = A^3/27$. }
  \label{fig:ZF_DAW}
\end{figure}

Equation (\ref{eq:ZF_KBM}) is derived in the $k_{\parallel0}=0$ limit, with the two terms in the square bracket  corresponding to ZF generation by DAW self-beating and radial envelope modulation, and can be used for studying the condition for ZF generation by KBM. \footnote{It is noteworthy that, the  KBM dispersion relation and polarization  can be sensitive to plasma parameters including $\omega_*$ and $\eta=L_n/L_T$. KBMs may have finite $k_{\parallel}$  in certain parameter regimes, as carefully investigated in Ref. \citenum{RMaPPCF2022}. In this sense, the investigation here can be more straightforwardly applied to BAEs.  However,  in this work  we will still take the $|k_{\parallel}V_A/\omega|\ll1$ limit for KBM, while leaving the more detailed analysis for a separate publication.}   It is worth noting that, in the uniform plasma limit, it will recover the results of ZF excitation by BAE \cite{ZQiuNF2016}.  Equation (\ref{eq:ZF_KBM}) can be re-written as
\begin{eqnarray}
\gamma^3_Z + A\gamma_Z +iB=0,\label{eq:cubic}
\end{eqnarray}
with the coefficients given by
\begin{eqnarray}
A &\equiv& \Delta^2 - \left(\frac{c}{B_0}k_Zk_{\theta}\right)^2\frac{|A_0|^2\hat{\beta}}{b_+\omega_0\chi_{iZ}}\frac{2}{\partial\epsilon_{A0}/\partial\omega_0}\nonumber\\
&&\hspace*{2em}\times \left\langle (J^2_+-J^2_0) \left(1-\frac{\omega^t_{*i}}{\omega}\right)_0\frac{F_0}{n_0}\right\rangle,\nonumber\\
B&\equiv&  \left(\frac{c}{B_0}k_Zk_{\theta}\right)^2\frac{|A_0|^2\hat{\beta}}{b_+\omega_0\chi_{iZ}}\frac{2}{\partial\epsilon_{A0}/\partial\omega_0}\nonumber\\
&&\times \left\langle (J_+J_ZJ_0-J^2_+)\left(\frac{\omega^t_{*i}}{\omega}\right)_0\frac{F_0}{n_0}\right\rangle.\nonumber
\end{eqnarray}
The conditions for ZF excitation by KBM, can thus be derived, by noting the expression of the pump KBM dispersion relation $\epsilon_{A0}$.   For the special case with  $B=0$, it is straightforward to see that ZF can   be driven unstable, as the nonlinear drive overcomes the threshold due to frequency mismatch. 
From the general  case with $B\neq0$, the condition for equation (\ref{eq:cubic}) to have a root with positive real part, can be determined from the    properties of cubic equations with one variable.  We, however, will only illustrate briefly the results from numerical solution of equation (\ref{eq:cubic}) in Fig. \ref{fig:ZF_DAW}. In Fig. 1(a), the dependence of the real part of $\gamma_Z$ with the  biggest real part on parameters $A$ and $B$ is given, and the regions for ZFZS exponentially growing is shown by the white region of Fig. 1(b), while the black region corresponds to ZFZS marginally stable with $\mbox{Re}(\gamma_Z)=0$. It is worth noting that, the boundary separating  the white and black regions, is given by $B^2/4=A^3/27$, from the properties of cubic equations. 
The  detailed analysis, however, will need more careful investigation with realistic plasma parameters and global KBM dispersion relation, and is, beyond the scope of the present work to formulate the general dispersion relation for ZFZS generation by DAW. The detailed analysis of ZFZS generation by KBM, of particular interest for turbulence transport in reactor scale tokamaks with plasma to magnetic pressure ratio significantly higher than present day machines, will be reported in a future publication.

\section{Summary and Discussion}
\label{sec:summary}

In this work, general equations  for zero frequency zonal structure (ZFZS) nonlinear excitation by drift Alfv\'en waves (DAWs) are derived, with contribution  of plasma nonuniformity and kinetic effects accounted for on the same footing.  It is found that, the   finite coupling between DAWs to effectively generate ZFZS, may from the radial modulation as the DAWs having different radial wavenumber, and self-beating where no difference of radial wavenumber  is required, corresponding to spontaneous excitation  and beat-driven  \cite{LChenPoP2024}, respectively.  This can be clearly seen from equation (\ref{eq:ZC}) for zonal current (ZC) generation, and equation (\ref{eq:ZF_KBM}) for zonal flow generation.  

The obtained nonlinear dispersion relation can, thus, be applied to study ZFZS generation by DAWs covering a broad frequency range.   In the first application, it is shown that the general dispersion relation can recover that of ZFZS excitation by TAE as effects associated with  $\omega_*$ is neglected \cite{LChenPRL2012}. For the modes in the TAE frequency range and ZC generation dominant, it is shown that, plasma nonuniformity will quantitatively modify the ZFZS generation process, while the qualitative picture is not changed.  For DAWs in the KBM frequency range with frequency comparable to diamagnetic frequency, where effect  of plasma nonuniformity is expected to be crucial, it is found that, the contribution of self-beating and radial envelope modulation renders the final nonlinear dispersion relation into a cubic equation of ZF growth rate, which will yield parameter regions for ZF excitation and marginally stable.  The detailed analysis with realistic KBM dispersion relation, however, is beyond the scope of the present work to formulate the general dispersion relation for ZFZS generation by DAW instabilities, and will be reported in a future publication.

\section*{Acknowledgement}
This work was  supported by the Strategic Priority Research Program of Chinese Academy of Sciences under Grant No. XDB0790000,  the National Science Foundation of China under Grant Nos. 12275236 and 12261131622, and  Italian Ministry for Foreign Affairs and International Cooperation Project under Grant  No. CN23GR02. The authors acknowledge   Dr. Fulvio Zonca (CNPS-ENEA and ZJU)  for fruitful discussions.


\end{document}